\documentclass[sort&compress]{elsarticle}
\usepackage{amssymb,amsmath}
\usepackage{units}

\usepackage{hyperref}

\newcommand{\e}{\mathrm{e}}
\renewcommand{\l}{\left(}
\renewcommand{\r}{\right)}
\newcommand{\q}{\frac{k}{a_0}}

\begin{document}

\title{Distinguishing between\\
  $R^2$-inflation and Higgs-inflation}

\author[lmu,inr]{F.~L.~Bezrukov}
\ead{Fedor.Bezrukov@physik.uni-muenchen.de}
\author[inr]{D.~S.~Gorbunov}
\ead{gorby@ms2.inr.ac.ru}

\address[lmu]{Arnold Sommerfeld Center for Theoretical Physics,
  Department f\"ur Physik,\\
  Ludwig-Maximilians-Universit\"{a}t, Theresienstr.~37, 80333
  M\"{u}nchen, Germany}
\address[inr]{Institute for Nuclear Research of the Russian Academy of
  Sciences,\\
  60th October Anniversary prospect 7a, Moscow 117312, Russia}
\date{November 18, 2011}

\begin{abstract}
  We present three features which can be used to distinguish the
  $R^2$-inflation Higgs-inflation from with ongoing, upcoming and
  planned experiments, assuming no new physics (apart form sterile
  neutrinos) up to inflationary scale.  (i) Slightly different tilt of
  the scalar perturbation spectrum $n_s$ and ratio $r$ of
  scalar-to-tensor perturbation amplitudes.  (ii) Gravity waves
  produced within $R^2$-model by collapsing, merging and evaporating
  scalaron clumps formed in the post-inflationary Universe.  (iii)
  Different ranges of the possible Standard Model Higgs boson masses,
  where the electroweak vacuum remains stable while the Universe
  evolves after inflation.  Specifically, in the $R^2$-model Higgs
  boson can be as light as 116 GeV.  These effects mainly rely on the
  lower reheating temperature in the $R^2$-inflation.
\end{abstract}

\maketitle

Early time inflation is a very attractive idea allowing to solve many
serious problems of the Hot Big Bang cosmological model originating in
the mystery of the initial conditions of our Universe, see
e.g.~\cite{Gorbunov:2011zzc}.  The inflation can be arranged with a
specific dynamics of only one degree of freedom---a scalar field.
Remarkably, many simplest models give different predictions for the
tilt of the scalar perturbation spectrum $n_s$ and for the ratio $r$
of squared amplitudes of tensor and scalar perturbations, and thus can
be distinguished experimentally by CMB observations.  On the contrary,
if predictions for $(n_s,r)$ coincide, it will be generally difficult
to determine which model is realized in Nature, if the measured value
of $(n_s,r)$ is close to the prediction.

One could expect this to be the situation for a pair of two minimal
inflationary models: $R^2$-inflation and
Higgs-inflation~\cite{Bezrukov:2007ep}.  They are \emph{minimal,}
because in order to solve the well-known problems of the Hot Big Bang
model they introduce very little of new physics.  Indeed, in the first
model with nonlinear modification of the Einstein--Hilbert action,
\emph{one and the same interaction---gravity---} takes care of
\emph{both inflation and subsequent reheating of the Universe.}  Only
one new degree of freedom---a scalar in the gravity sector---emerges
and only one new parameter (in front of the $R^2$-term) is present.
In the second model \emph{no new degrees of freedom} appear: thanks to
the non-minimal coupling to gravity (with only one new corresponding
parameter) the Standard Model (SM) Higgs boson plays the role of
inflaton and its couplings to the SM fields are responsible for
subsequent reheating.\footnote{Let us note, that introduction of
  additional light scalar degree of freedom (or several) also leads to
  a plethora of good models, see
  e.g.~\cite{Bezrukov:2009yw,Lorenz:2008je}.  However they are
  naturally leading to obviously very different inflationary
  parameters, and are less minimal, so we do not consider them in this
  Letter.}  Remarkably, these minimal inflationary models \emph{can be
  augmented} with a small amount of new interaction terms and degrees
of freedom to accomodate all currently firmly established experimental
evidences of beyond the SM physics (neutrino oscillations, Dark
Matter, and baryon asymmetry of the Universe), see
e.g.~\cite{Gorbunov:2010bn,Gorbunov:2012ij} and
\cite{Bezrukov:2008ut,Bezrukov:2011sz}.  These modifications do not
interfere with the inflationary and reheating dynamics, and hence are
irrelevant for the present study.  A particular example of the
renormalizable model is $\nu$MSM \cite{Asaka:2005an,Asaka:2005pn},
which is a SM extension with three right handed neutrinos capable of
explaining simultaneously the active neutrino masses (by see-saw like
mechanism), DM (sterile neutrino at keV scale) and baryon asymmetry of the
Universe (by a specific variant of leptogenesis), for a review see
\cite{Boyarsky:2009ix}.  Other examples may also be suggested,
e.g.~\cite{Gorbunov:2010bn,Gorbunov:2012ij,Bezrukov:2011sz}.  We
stress, however, that to make exact predictions for the inflationary
parameters the evolution of the Universe needs to be known both during
and after the inflation.  This is certainly achieved in any extension,
which does not introduce new scales between the electroweak and
inflationary scales, and this is true in the $\nu$MSM.

Let us return to the inflationary models proper.  The similarity
between the two models is clearly seen in the Einstein frame, where
the potentials for canonically normalized inflatons $\chi$ (scalaron
and the Higgs boson, respectively) are identical for the large fields
$\chi\gg U^2/M_P$, relevant for inflationary dynamics,
\begin{equation}
  \label{common-potential}
  V(\chi) = \frac{U^4}{4}\left(
    1-\exp \l -\frac{2\chi}{\sqrt{6}M_P}\r
  \right)^{2}
  .
\end{equation}
Here $M_P\equiv 1/\sqrt{8\pi G_N}=\unit[2.44\times10^{18}]{GeV}$ is
the reduced Planck mass, and $U$ is the only parameter controlling the
scale of the potential.  For $\chi\gtrsim M_P$ the slow roll
conditions are satisfied and inflationary stage takes place.  The
expressions of the constant $U$ in terms of the fundamental constants
of the theory ($\xi$ and $\lambda$ for the Higgs-inflation
\cite{Bezrukov:2007ep} and $\mu$ for the $R^2$-inflation
\cite{Gorbunov:2010bn}) are different in the two models, but as far as
its value $U=M_P/\sqrt{47000}$ is uniquely defined by the
normalization of the amplitude of the primordial density
perturbations, both models seem to predict the same set of $(n_s,r)$.

Are there any differences between these two models, which can be
resolved at the present level of experimental techniques?  The answer
is positive and we discuss the details in this \emph{Letter.}


Certainly, the two models are different.  $R^2$-inflation modifies the
gravitational sector and exploits as the inflaton one new degree of
freedom---scalaron---in addition to the SM
and hence to the Higgs-inflation.  The
Higgs-inflation modifies the interaction of gravity with only one of
the existing particles---the Higgs boson---making the Higgs field
itself the inflaton.

The key observation is that right after inflation, starting from the
onset of inflaton homogeneous oscillations, interactions of the two
inflatons with the SM fields are absolutely different.  Being one of
the gravitational excitations the scalaron, similarly to the graviton,
interacts with gravitational strength, i.e.\ all scalaron coupling
constants are suppressed by an appropriate power of the Planck mass.
The Higgs boson interacts with the weak bosons and top quarks with
couplings of order one.  This circumstance drastically changes the
post-inflationary evolution of the Universe, which takes place between
inflation and the hot stage (preheating).  Indeed, though in both
models at this stage the inflaton exhibits free-field oscillations and
the Universe is in a matter dominated stage, the inflaton couplings to
the other fields play a crucial role.

In the case of Higgs-inflation one can see, 
that while at high scales the Higgs boson 
is nearly decoupled from all other SM
fields, at the energy scale of the order $U^2/M_P\sim\unit[10^{13}]{GeV}$ its
interaction regain the SM form~\cite{Bezrukov:2007ep}.  
So, when the oscillation amplitude of
the field after inflation drops below this value, the energy is
effectively transferred into all SM degrees of freedom.  The detailed
analysis \cite{Bezrukov:2008ut,Bezrukov:2011sz} of the field decay
during the matter dominated stage shows that the Higgs-inflaton field
rapidly produces weak bosons, which subsequently decay into all other
SM particles and reheat the Universe, leading to a slightly higher
than $U^2/M_P$ temperature
\begin{equation}
  \label{TrH}
  T^{\mathrm{reh}}_H \simeq \unit[6\times 10^{13}]{GeV},
\end{equation}
with uncertainty factor about two.

In the $R^2$-inflation the scalaron coupling to all fields is Planck
scale suppressed, and the reheating mainly occurs via its decays into
the SM Higgs bosons \cite{Starobinsky:1980te,Vilenkin:1985md}, which
immediately rescatter into SM particles 
(scalaron couplings to
all other fields are additionally suppressed due to conformal
symmetry).  The reheating temperature in the model is significantly
lower \cite{Gorbunov:2010bn},
\begin{equation}
  \label{TrR2}
  T^{\mathrm{reh}}_{R^2} = \unit[3.1\times 10^9]{GeV}.
\end{equation}

The difference in reheating temperatures yields three important
consequences each providing with features (potentially) experimentally
observable and allowing for discriminating between the models.  These
consequences are slightly different numbers of e-foldings before the
end of inflation for the moment of the horizon exit of the same pivot
scale of WMAP, collapses of small scale scalaron perturbations in
$R^2$-model in the post-inflationary Universe and different regions of
the SM parameter space, where the electroweak vacuum remains
sufficiently stable today as well as in the very early Universe.  Let
us discuss these issues in turn.

\paragraph{Different e-folding numbers}
The post-inflationary history of the Universe differs in the two
models: the pre-Big-Bang matter dominated stage lasts much longer in
$R^2$-inflation.  As a result, the matter perturbations of a given
wave-length at present (including the scale used for normalization of
WMAP) were of different sizes at the time of horizon crossing at the
inflationary stage in these two models.  Indeed, the wave-length
scales as the scale factor, $\propto a(t)$, when the Universe expands,
and at matter-domination and radiation-domination the scale factor
grows differently (as $a(t)\propto t^{2/3}$ and $a(t)\propto t^{1/2}$,
respectively).  Hence, at the beginning of the post-inflationary stage
the wave lengths of a perturbation of a given present scale were
different in these two cosmological models.  At the end of inflation
their wave-length exceeded the horizon size.  Their amplitudes have
been frozen earlier at the inflationary stage when the wave-length of
the stretched perturbations crossed the horizon (see details in
e.g.~\cite{Gorbunov:2011zzc}).  Thus, the e-folding numbers (and,
respectively, the field values in~(\ref{common-potential}),
corresponding to this moment of time) of the horizon crossing for the
modes of the presently observed pivot scale are different in the two
inflationary models.

Let us calculate the number of e-foldings for the fluctuations of
conformal momentum $k$.  We are interested in fluctuations which
physical momentum today corresponds to the WMAP pivot scale
$k/a_0=0.002/\unit{Mpc}$, where $a_0$ is the present scale factor
(below we use subscript `0' to denote the present values of
parameters).  At inflationary stage the fluctuations exit the horizon
when the physical momentum $k/a$ drops below the value of the Hubble
parameter $H=\dot{a}/a$ determining the expansion rate.  Marking the
values of all parameters at the horizon crossing with asterisk, we
write
\begin{equation*}
  H_*=\frac{k}{a_*}=\q\frac{a_0}{a_r}\frac{a_r}{a_e}\frac{a_e}{a_*}
  \equiv\q\frac{a_0}{a_r}\frac{a_r}{a_e}\e^{N},
\end{equation*}
hereafter subscripts `e' and `r' refer to the values of parameters at
the end of inflation and at the end of reheating, respectively.  The
change of the scale factor after reheating is
\begin{equation*}
  \frac{a_r}{a_0}=\left(\frac{g_0}{g_r}\right)^{1/3}\frac{T_0}{T_r},
\end{equation*}
where $g_r$ is the number of degrees of freedom (d.o.f.) in the
primordial plasma at reheating and
$g_0=2+\frac{7}{8}\cdot2\cdot3\cdot\frac{4}{11}$ is the present
effective number of relativistic d.o.f.\ taking into account different
neutrino temperature.  Since both models exhibit matter dominated
expansion between inflation and reheating, we get for the change of
the scale factor during preheating
\begin{equation*}
  \frac{a_r}{a_e} =
  \left(\frac{V_e}{g_{r}\frac{\pi^2}{30}T_r^4}\right)^{1/3}.
\end{equation*}
Collecting everything and using the Friedman equation for the Hubble
parameter $H=\l V_e/(3M_P^2)\r^{1/2}$, we get
\begin{multline}
  \label{N-formula}
  N = \frac{1}{3}\log \l \frac{\pi^2}{30\sqrt{27}}\r
  -\log\frac{(k/a_0)}{T_0g_0^{1/3}}
  +\log\frac{V_*^{1/2}}{V_e^{1/4}M_P}
  -\frac{1}{3}\log\frac{V_e^{1/4}}{\unit[10^{13}]{GeV}}
  \\
  -\frac{1}{3}\log\frac{\unit[10^{13}]{GeV}}{T_r} ,
\end{multline}
The first term in \eqref{N-formula} contains model-independent
numbers, the rest terms in the same line vary with the change of the
moments of horizon crossing and end of inflation very mildly
(sub-logarithmically),
\[
  V_*\approx\frac{U^4}{4},\qquad
  V_e\approx\frac{U^4}{4(1+\sqrt{3/4})^2}.
\]  
The main difference in $N_*$ comes from the last term in
\eqref{N-formula}, so that approximately 
\[
  N_*\approx 57-\frac{1}{3}\log\frac{\unit[10^{13}]{GeV}}{T_r}.
\]
For the models under discussion with the reheating temperatures
(\ref{TrH}) and (\ref{TrR2}) one obtains numerically from
(\ref{N-formula})
\[
  N_{H}=57.66,\qquad N_{R^2}=54.37.
\]
This discussion is applicable to both scalar (inflaton) and tensor
(graviton) perturbations.  The different sizes of horizon imply
different values of the inflaton potential \eqref{common-potential}
and hence different values of the inflaton field $\chi_*$.  In this
way we finally arrive at different values of parameters of scalar and
tensor perturbations (see details in e.g.~\cite{Gorbunov:2011zzc}),
which for the present models sharing the same potential
\eqref{common-potential} at inflationary stage are mainly (up to
corrections $O(U/M_P)$) determined by the number of e-foldings, i.e.\
$n_s\simeq1-8(4N+9)/(4N+3)^2$ and $r\simeq192/(4N+3)^2$
\cite{Bezrukov:2008ut}.  Using exact formulas (see
e.g.~\cite{Gorbunov:2011zzc,Bezrukov:2008ut}) gives numerically
\begin{align*}
  \text{Higgs-inflation: } & n_s=0.967,\quad r=0.0032,\\
  R^2\text{-inflation: }   & n_s=0.965,\quad r=0.0036.
\end{align*}
The difference is small, at the level of $10^{-3}$, but such an
accuracy is close to achievable at Planck experiment (expected
precision for $n_s$ is $0.0045$, see \cite{PlanckBlueBook}) and CMBPol
experiment (precision for $r$ is $10^{-3}$ or even $0.5\times10^{-3}$,
and for $n_s$ up to $0.0016$ \cite{Baumann:2008aq}).  Note that an
additional test for all large-field inflationary models could become
possible if the tilt of the tensor fluctuations spectrum $n_T\approx
-r/8$ is measured.

The primordial tensor perturbations can be also directly detected as
gravity waves by the advanced stages of DECIGO project \cite{DECIGO}.
Note also, that the matter dominated stage causes reduction of the
gravity waves amplitude for high frequency modes.  For the reheating
temperature of $T_r\sim\unit[10^9]{GeV}$ the gravity waves are
suppressed at frequencies above approximately $\unit[10]{Hz}$
\cite{Kuroyanagi:2011fy} (and above lower frequencies for lower
reheating temperatures).  While being not testable by the currently
proposed detectors, the value is not far from explorable region.

\paragraph{Gravitational waves at matter dominated stage}
When the matter perturbations we discussed above enter the horizon,
they grow proportionally to the scale factor at the matter-dominated
stage.  The observed CMB anisotropy and large scale structure can be
explained with primordial (inflaton) energy density perturbations at
the level $\delta \rho/\rho\sim 10^{-5}$.  With long post-inflationary
matter dominated stage in $R^2$-inflation certain small scale
perturbations have enough time to enter horizon and grow up to
$\delta\rho/\rho\sim 1$ evolving towards the nonlinear stage.  Then
one expects production of small scale self-gravitating structures from
the scalaron condensate, which does not change the reheating
\cite{Gorbunov:2010bn}, but can give rise to gravity waves emission by
inflaton clumps.

Indeed, at the late post-inflationary stage the scalaron behaves like
homogeneously oscillating free scalar field (inflaton), which was
argued \cite{Jedamzik:2010hq,Schutz:1984nf} to be capable of producing
short-length gravity waves due to collapses of inflaton perturbations,
merging of inflaton clumps (halos), final evaporation of inflaton
clumps (halos) during reheating (scalaron decays in the case of
$R^2$-inflation).  The signal from the latter processes falls
\cite{Jedamzik:2010hq} right in the region to be probed at the
advanced stage of DECIGO project \cite{DECIGO} on gravity waves
measurements.  This allows for an independent test of the
$R^2$-inflationary model.

\paragraph{Allowed Higgs boson masses}
As far as in both models there are no new physics between electroweak
and inflationary scales, one can expect bounds on the Higgs mass from
the absence of strong coupling and stability of the electroweak
vacuum.

The models have to be in the weak coupling regime for all the SM
couplings up to the inflationary scale, which is determined by value
of the Hubble parameter at the end of inflation,
$H_{e}\sim\unit[10^{13}]{GeV}$.  In particular, large self-coupling of
the Higgs boson is forbidden, if the corresponding Landau pole is at a
lower energy scale.  This places an upper limit on the Higgs boson
mass (proportional to the square root of the self-coupling) in both
models at the level
\cite{Bezrukov:2009db,Maiani:1977cg,Cabibbo:1979ay,Lindner:1985uk,Hambye:1996wb}
\[
  m_h\lesssim \unit[194]{GeV}.
\]
At present this bound is superseded by the direct searches at LHC,
which give a stronger upper limit $m_h\lesssim\unit[146]{GeV}$
\cite{CMS-HIG-23}.

The lower limit on the Higgs self-coupling comes from the requirement
of a sufficient stability of the electroweak vacuum in the SM.  At a
small self-coupling (corresponding to $m_h\lesssim\unit[129]{GeV}$,
see
\cite{Bezrukov:2009db,Maiani:1977cg,Cabibbo:1979ay,Lindner:1985uk,Hambye:1996wb})
the vacuum becomes unstable at large energy scales, mostly due to the
top-quark corrections.  The true vacuum in this case appears at a very
large value of the Higgs field.  The absolute stability of the
electroweak vacuum is not required, because the decay rate of our
vacuum (at zero temperature) is very small, and it can survive for the
time equal to the age of the Universe for any
$m_h\gtrsim\unit[111]{GeV}$ \cite{Espinosa:2007qp}, which is below the
LEP bound.

However, in the early Universe (right after inflation, at the early
matter dominated stage, or later at the hot stage) the Higgs field may
be caught in this vacuum.  As far as this is not the case for the
observed Universe, this implies a lower limit on the Higgs
self-coupling and hence a lower limit on the Higgs boson mass.

For the case of the $R^2$-inflation the Higgs field does not evolve to
the large field minimum during inflation, if the inflation started
with a reasonable small value of the Higgs field.  At the same time,
the metastable vacuum can decay at high temperature right after
preheating.  The corresponding bound for the reheating temperature
(\ref{TrR2}) can be obtained from Ref.~\cite{Espinosa:2007qp},
\begin{equation*}
  m_h^{R^2} > \left[
    116.5
    + \frac{m_t-\unit[172.9]{GeV}}{1.1} \times 2.6
    - \frac{\alpha_s(M_Z)-0.1181}{0.0007} \times 0.5
  \right] \unit{GeV}
  .
\end{equation*}
The limit depends on the top-quark mass $m_t$, strong coupling
constant $\alpha_s$, and also has systematic uncertainty of about
\unit[3]{GeV} from higher loop corrections.  Note, that for variations
of the $R^2$-inflation with lower reheating temperature the viable
interval of the Higgs boson mass may be even wider and almost overlap
with the direct lower bound from LEP2, $m_h>\unit[114.4]{GeV}$
\cite{Barate:2003sz}.

For the Higgs-inflation the lower limit from the thermal decay with
reheating temperature (\ref{TrH}) is about $m_h>\unit[120]{GeV}$.  But
the bound in this case is even stronger, because at the end of
inflation the model directly ends up with the large Higgs field value
(of the order at least $\unit[10^{13}]{GeV}$).  So, if the SM
potential has a minimum at that scale the evolution stops there even
before reheating, leading to the bound \cite{Bezrukov:2009db} (see
also
\cite{Bezrukov:2008ej,DeSimone:2008ei,Barvinsky:2009fy,Barvinsky:2009ii})
\begin{equation*}
  \label{2loop}
  m_h^{\mathrm{H}} > \left[
    129.0
    + \frac{m_t-\unit[172.9]{GeV}}{1.1} \times 2.1
    - \frac{\alpha_s(M_Z)-0.1181}{0.0007} \times 0.5
  \right] \unit{GeV},
\end{equation*}
also with a systematic error of about $\unit[2]{GeV}$, related to
higher loop corrections.  Note, that this analysis depends on the
assumptions about the UV completion of the Higgs inflationary model
\cite{Bezrukov:2010jz}, specifically on the value of the non-minimal
coupling $\xi$ after inflation and the shape of the potential during
reheating.

\paragraph{Conclusions}
The models of Higgs-inflation and $R^2$-inflation, though providing
identical inflationary potential and thus exhibiting identical
dynamics at the inflationary stage, can still be distinguished
experimentally by cosmological observations.  The key point is that
both models, having no additional beyond the SM dynamics at high
energies except for the inflationary one, provide the full description
of the Universe evolution.  Specifically, they both allow to study the
reheating process, and have different reheating temperatures.  This
allows to determine exactly the predictions for the CMB parameters in
the models and gives rise to possible distinguishing signals in future
gravitational wave experiments.  Different reheating mechanisms also
lead to a much more constrained region of allowed Higgs masses for the
Higgs-inflation, which can be very soon fully explored at LHC.

F.B. would like to thank H.~M.~Lee for a fruitful discussion on the
variants of the Higgs-inflation model.  The work is supported in part
by the grant of the President of the Russian Federation
NS-5590.2012.2.  The work of F.B. is partially supported by the
Humboldt foundation.  The work of D.G. is partially supported by
Russian Foundation for Basic Research grants 11-02-01528-a and
11-02-92108-YAF\_a, and by the SCOPES.


\bibliography{R2-vs-Higgs}
\bibliographystyle{elsarticle-num}

\end{document}